\relax
\documentclass[11pt,letterpaper, twocolumn]{article} 
\usepackage{graphicx} 
\usepackage{graphicx}  
\usepackage{fancyhdr}
\title{On Interactive Machine Learning and the Potential of Cognitive Feedback}
\author{Chris J. Michael, Dina Acklin, \ and Jaelle Scheuerman\\
U.S. Naval Research Laboratory\\
Stennis Space Center, MS, U.S.A.
}
\date{Submitted to DMAIDA 22 December 2019\\
Accepted 31 January 2020}

\begin{document}

\maketitle
\pagestyle{fancy}
\thispagestyle{fancy}

\begin{abstract}
In order to increase productivity, capability, and data exploitation, 
numerous defense applications are experiencing an integration of 
state-of-the-art machine learning and AI into their architectures.
Especially for defense applications, having a human analyst in the loop
is of high interest due to quality control, accountability, and complex
subject matter expertise not readily automated or replicated by AI.
However, many applications are suffering from a very slow transition. 
This may be in large part due to lack of trust, usability, and productivity,
especially
when adapting to unforeseen classes and changes in mission context.
Interactive machine learning is a newly emerging field in which machine
learning implementations are trained, optimized, evaluated, and exploited
through an intuitive human-computer interface.  In this paper, we
introduce interactive machine learning and explain its advantages and
limitations within the context of defense applications.  Furthermore,
we address several of the shortcomings of interactive machine learning
by discussing how cognitive feedback may inform features, data, and results
in the state of the art.  We define the three techniques by which cognitive
feedback may be employed: self reporting, implicit cognitive feedback, and 
modeled cognitive feedback.  The advantages and disadvantages of each 
technique are discussed.
\end{abstract}

\section{The Emergence of Interactive Machine Learning}
The vast majority of modern-day research in 
machine learning presents algorithms and implementations
that do not consider human interaction.  For example, the flourishing field of
deep learning research is evaluated mainly by classification accuracy over
large curated datasets and generative models.  
This approach, referred to as Automatic Machine Learning (AML) or sometimes
conventional machine learning, forgoes the integration of dynamic human feedback
into the system.
Though undoubtedly useful for
commercial big-data problems, there are many scenarios -- especially in defense --
where applying AML falls short in practice.  
For instance, applications
at the tactical edge may suffer from smaller quantities of labeled examples for 
training.
Moreover, classifiers may struggle to adapt to changes in data context quickly enough
to be considered viable by an analyst, particularly in scenarios where the mission
demands quick turn-around time.
Many of these issues may be mitigated by emerging implementations of the {\it interactive
machine learning} (IML) paradigm, which capitalizes on human input in order to improve
machine learning implementations \cite{Fails2003}.  Unlike approaches that leverage
AML, IML implementations allow for 
classifiers to very quickly train and apply newly discovered information with the help
of a human subject-matter expert, which we refer to in this article as the analyst.

In general, IML may be
described as a machine learning implementation where one or more analysts
iteratively improve a model for automation by manipulating an interface that is
tightly coupled to the desired task at hand.  
There are four main
components to any IML implementation.  The first component is the data associated
with the task.  Examples of such data include remotely sensed imagery, textual
information such as reports, and spatiotemporal tracks of moving objects.
The second component, referred
to in this study as the {\it machine}, is the mathematical
model that tries to estimate or automate the desired task.  Ostensibly,
this can be seen as a black-box, but we will discuss
the properties of a successful IML classifier later in the article.
The third component of IML is the Human-Computer Interface (HCI).  The HCI may be
as conventional as software receiving input through a keyboard and mouse, 
which is what we assume in this article, or as
specialized as vehicle controls, immersive environments, and brain interfaces.
The application is designed to allow immediate and 
intuitive presentation of the machine's classification on a manageable set of data.
This data is then either confirmed or manipulated to be correct by the analyst, who
is the last but most
important component of an IML system.  In this article, we discuss IML within
the context of improving productivity and decision making for an analyst with a very
specific task that requires subject-matter expertise.
Though, as exemplified above, IML may be deployed in
a wide variety of ways, we feel that deployment in this context has the greatest
potential for impact in defense applications.  There are several studies that
provide excellent perspectives of the current state of the art in IML outside of this
scope \cite{Dudley2018,Wu2019,Robert2016}.

\begin{figure*}[t]
\centering
\includegraphics[width=0.7\textwidth]{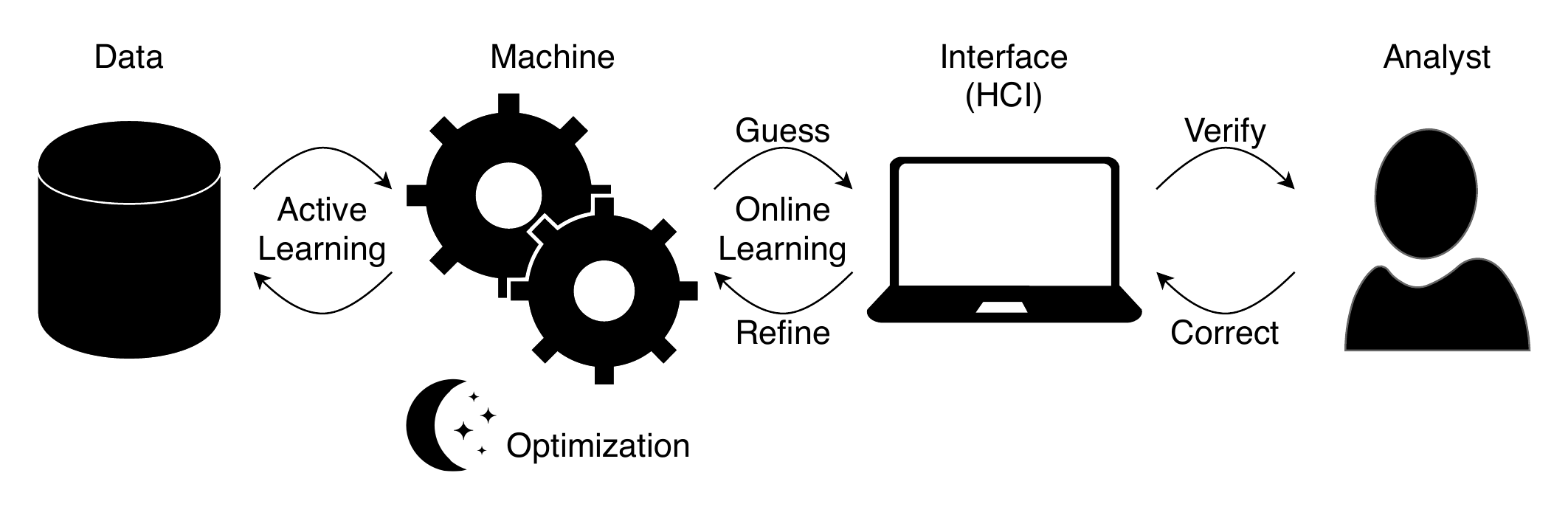} 
\caption{A common architecture for interactive machine learning implementations.}
\label{fig:iml}
\end{figure*}

A common architecture for IML implementations is shown in Figure \ref{fig:iml}.
The data on which the analyst must perform a task may either be completely 
available in a database or sequentially available as a stream.  {\it Active learning}
may be used to pull the most effective data points from this database for labeling,
as will be discussed in the next section.  A machine for predicting the
data is then used to present guesses for the task at hand to the analyst.  The
analyst must verify each of these guesses and correct any mistakes via the HCI.
Once the verification step is completed for the current iteration, the machine
will immediately learn from the corrections and/or confirmations.  The process will
then repeat by the machine gathering data examples and presenting guesses to the
user once again.  When the time comes for the analyst to leave their duty
station, the machine model may optimize on the data that has
been labeled in order to maximize its accuracy.  This way, the most effective
machine will be available once the analyst returns to duty.
It is important to note that the 
machine may be deployed as a centralized general-purpose classifier that combines
the work done by multiple analysts, or it may be deployed locally to be custimized
towards the individual analyst.

The focus of this article is to introduce IML within the context of analyst-driven
applications relevant to defense while highlighting research gaps, the most
important of which involves incorporation of cognitive feedback.
We choose not to discuss
manual
{\it model interactions} such as feature selection \cite{Raghavan2006} or model
selection \cite{Talbot2009}, which are processes whereby analysts directly optimize
machine models.
Rather, we choose to present implementations 
that can be used effectively by an analyst who is a subject
matter expert for the task at hand and not knowledgeable 
in machine learning or statistical theory.
Defense analysts hold invaluable
subject-matter expertise for the mission, and it is unreasonable to assume that
they must learn or worry about data-scientific concepts.
Because intuitive HCIs may be designed to be 
congruent to their task, IML has great potential to leverage the power
of modern-day ML while not burdening the analyst with parameter tuning, data curation,
or any of the other burdens implicit to AML.

The next section will describe three examples of IML implementations that highlight the
current state of the art.  The section that follows will
iterate through several advantages, shortcomings, and gaps in the state of the art.  In
the penultimate section, we specify the ways in which cognitive feedback 
may be used to address the 
shortcomings and gaps of IML with respect to defense applications.  Finally, we conclude
with commentary on prospects for future research.

\section{Interactive Machine Learning in Action}

In order to frame a more detailed discussion of IML, we now describe several IML 
implementations that have been presented in peer-reviewed literature.  We specifically
choose three application areas that are analyst-driven: region digitization,
textual translation, and video annotation.  These examples demonstrate the potential for
IML to improve both the machine performance and the user experience with autonomy.

Geographic region digitization is a highly demanded yet arduous task whereby regions
such as bodies of water and other land cover are digitized from remotely sensed
images, usually within a Geographic Information System \cite{Hossain2019}.
Once digitized, regions may be represented in mapping products for geospatial situational
awareness, climate-level studies, land surveys, and many other applications.
Although numerous AML approaches to region digitization have been presented
in the literature, they are not widely 
adopted in practice.  This is most likely due to the all-or-nothing yield of AML 
approaches: If the machine incorrectly digitizes a region, 
it may be more burdensome for an analyst to correct than to start from scratch.
Therefore, an analyst may prefer
to digitize manually to circumvent frustration and presumably lower their workload.
In order to address these shortcomings, an IML implementation for region digitization,
named the Geospatial Region Application Interface Toolkit (GRAIT),
is presented as a human-machine team application \cite{Michael2019}.
The authors address the all-or-nothing approach
to region digitization with an IML implementation where a region is digitized iteratively.
In each iteration, the machine guesses the placement of a certain number of 
vertices of the contour
and presents them to the analyst for verification.  For each vertex presented,
the analyst may either correct its placement by clicking and dragging it to an appropriate
location or simply confirm its correct placement by not interacting with it.
The analyst indicates via button press when all the
vertices of the current iteration are corrected or confirmed.  The machine will then train
on the finalized vertex locations, and the process will continue until the region is
completely
digitized.  In order to prevent inducing too high of a cognitive load on the analyst,
an uncertainty model is used to estimate the probability of incorrect vertex placement
and limit each iteration to around 2 incorrectly placed vertices.  Results
show that with no prior training data, the IML implementation accurately places 84\% of
vertices correctly in 4 separate image sets of 4 images each.

Another area where IML approaches show promise is that of textual language translation,
commonly referred to as {\it machine translation}.  While bodies of work
 in this field attempt to
replace human translators with machine models, many of which are AML implementations
\cite{Koehn2009}, the current state of the art is far from perfect.
As with region digitization, fully-automatic approaches may hinder rather than help the 
performance of a translator at times when too many mistranslated words may induce excessive
cognitive load.  Because of this, many approaches to machine 
translation are realized through a human-machine team.  
An IML approach to machine translation 
aims to remedy these issues by implementing iterative learning and modeling the
informativeness of each machine translation at a fine-grained level
\cite{Gonzalez2014}.  In this approach,
an initial guess of a sentence translation is given to the user based on a metric
of informativeness.
The user will then make corrections to the guess by changing the first incorrect
letter of the translation.  The machine in turn suggests a new translation 
under this assumption.
This process continues, with the machine immediately training on corrected data for
future translations.  Results
show that employing this IML-based method produces twice the translation
quality, a metric specific to machine translation, per user interaction over AML approaches.

Lastly, IML implementations have emerged for the difficult task of video annotations, where
the amount of data generated per day has far surpassed the ability of analysts to inspect.
When successful, annotated video allows for critical advantages such as the ability to search
for events, quantify behavioral analytics, and study natural phenomena.
Though many AML approaches to
video analytics exist, they are typically tied to certain 
certain features of interest within some constrained context \cite{Anan2017}.
In cases where context may
change and the features of interest are unknown, AML implementations for automatic video
annotation may be rendered incorrect or infeasible.
This is especially true
in cases where context has changed or features of interest are unknown beforehand.
An IML implementation of video 
annotation named
Janelia Automatic Animal Behavior Annotator (JAABA) demonstrates a semi-automatic approach
to assess animal behavior \cite{Kabra2013}.
JAABA allows for a user to annotate a video frame 
with an arbitrary label, for instance {\it jump}.
Then, using trajectory information extracted from the video,
the machine trains on the given label and presents classification results both
 at the level
of the current video and a database of numerous animal videos.  The machine also provides
confidence levels for each classification to guide further labeling by the user.  This
process is repeated iteratively until an ideal classifier is attained.  JAABA was used
to create the first ML-driven behavior classifier over a diverse set of animals.

With these three examples in mind, a more detailed explanation of the advantages,
limitations, and gaps of IML will follow.

\section{Advantages, Shortcomings, and Gaps}

\subsection{Advantages}
The advantages of IML approaches directly address
many of the shortcomings that defense applications exhibit when utilizing ML.
Numerous defense applications suffer from a shortage of labeled training
examples due to a lack of crowd sourcing and the ever changing state of platform
technologies among other reasons.  As such, deep models relying on large amounts of
labeled examples cannot be adequately trained.
IML addresses the shortage of training data
by providing an interface that allows for incorrect classifications to be
immediately corrected and integrated into the machine model.  In fact, 
several IML
implementations may work well with no prior labeled data, which is usually referred to
as the {\it cold start problem} \cite{Lika2014}.
Additionally, the HCI
allows for correction through an intuitive interface that potentially reduces
the burden of data labeling.  This allows an analyst to leverage their current
subject-matter expertise -- that of the application and data context -- and 
circumvents the need
to play the role of data scientist.  

Defense problems must be very adaptable to context changes from one region of interest
to the next. 
In order to accommodate this, any autonomy must immediately adapt to 
such changes at the pace of the analyst.  Therefore, IML implementations typically
apply active learning and online learning techniques in order to improve effectiveness.
Active learning research entails the study of uncertainty or similarity metrics in 
order to develop a mathematical understanding of the likelihood that a machine will
classify future data points correctly \cite{Quionero2009}.
The field
of online machine learning involves models that may train in stride to adapt to new
situations quickly while optimizing exploration vs. exploitation \cite{Bottou1998}.

Problems related to defense must sometimes be deployed at the tactical edge.  In such
situations, computational resources and downtime may be scarce.  IML directly
addresses this problem, since most IML implementations are meant to be deployed on
desktop computers.  
In all three examples of IML presented in the
previous section, online and active learning strategies are employed to 
iteratively build high-performance classifiers.  Active learning is also used to
gage the load of examples presented to the user, both by correlating uncertainty
to the probability of an incorrect classification and by providing a priority for
the analyst to manage their own work flow.  Both HAML and
JAABA support cold-start cases.

\subsection{Shortcomings}

Perhaps the most obvious shortcoming if IML is that the HCI and machine implementation
must be tightly coupled to a specific application.  This entails much more effort
in the development of applications, since they must be built and studied uniquely
towards an explicit work flow.  This differs greatly from
AML approaches, where for the most part implementations are general-purpose and
specificity is implied through parameterization and classes for labeling.  Studies
define a general-purpose methodology for HCI, but this
research is young and remains mostly theoretical in nature \cite{Martinez2019}.

Deep models of machine learning exhibit very impressive results relating to throughput
of data and classification times.  IML implementations currently lag behind in these
results.  This is in part due to the nature of online machine learning; namely, the need
to have tight classification and training cycles.  However, research is trending
more towards online and active learning problems, and IML-inspired classifiers with
competent performance are emerging \cite{Langford2007,Lu2013}.

A further issue with IML is that overfitting may occur more frequently
since data is generally labeled iteratively.
Overfitting occurs when prior training data causes the model
to correlate too tightly to features that do not justify the desired outcome.  
For example, one of the geographic sites in the
GRAIT study is Johnson Lake, WA.  The first three images show the shoreline in roughly
the same location.  The fourth image shows the lake with a receded shoreline.  Though the
shoreline may be spotted by an analyst clearly in the fourth image, the classifier overfit
to spaital features and thus incorrectly identified the shoreline.  This also caused the
uncertainty calculations for the image to be undershot.  AML approaches to overfitting
typically require optimizing machine parameters or adding diversity to datasets, both of 
which typically require large amounts of computation and thus long turnaround times not
conducive to successful IML implementations.  Therefore, reinforcement meta-learning,
whereby active learning implementations are informed
by corrections via specialized ML implementations, may be employed 
to adapt quickly to situations
where overfitting is inevitable \cite{Bachman2017}.

\subsection{The Cognitive Gap}

Although frequently mentioned as a future direction of study,
perhaps the largest identified gap in IML research is the lack of 
formalization and quantification of cognitive implications
from the analyst.  For instance, the IML machine translation
study \cite{Gonzalez2014} mentions specifically that the applied 
technique lessens the cognitive load of the translator by utilizing cost-sensitive metrics
such as informativeness.  However, the study does not perform any human-factors research
to back support this claim, though it is mentioned as future work.
As another example, the study presenting GRAIT uses mathematically modeled uncertainty calculations
to meter the workload at each iteration.
Though it is shown statistically that these uncertainty calculations
correlate to the probability a vertex is placed correctly, results focus more on vertex placement
accuracy and do not consider multiple load levels (e.g. the number of expected incorrect vertices
is set to two for the entire study).  Human factors research is also slated as future work.
Both of these studies appreciate that there must be thresholds of cognitive load taken into account
by the IML system for a successful implementation, but it is apparent that human-factors research
is inevitable.

\section{The Implications of Cognitive Feedback}

Due to its interactive nature, IML most certainly is a human-in-the-loop endeavor.  Several studies
have highlighted difficulties that may arise from trust, safety, and quality
\cite{Dudley2018,Groce2013,Gillies2016,Turchetta2019}.  This section is devoted to discussing
the potential of researching and integrating models of human cognition as feedback for IML, which
is not often mentioned in the state of the art.  We also make the argument that cognitive feedback
directly addresses the shortcomings of IML. The topic of cognitive feedback is especially
useful for defense-related problems, where trust, safety, and quality of ML implementations is a
prerequisite for adoption.  Without analyst-driven cognitive feedback, an IML system can very
quickly fall flat, which is illustrated in the following region digitization example.

Consider the analyst using GRAIT to digitize the fourth image of Johnson Lake as explained in
the previous section.  Recall that the machine is overfit, and thus its model for uncertainty is 
undershot.  Because of this, the machine places
10 vertices, 8 of which are incorrectly placed.  If the analyst continues, they
will spend more time correcting the misplaced vertices than manually digitizing the lake without
the help of the machine. 

\begin{figure*}[t]
\centering
\includegraphics[width=0.9\textwidth]{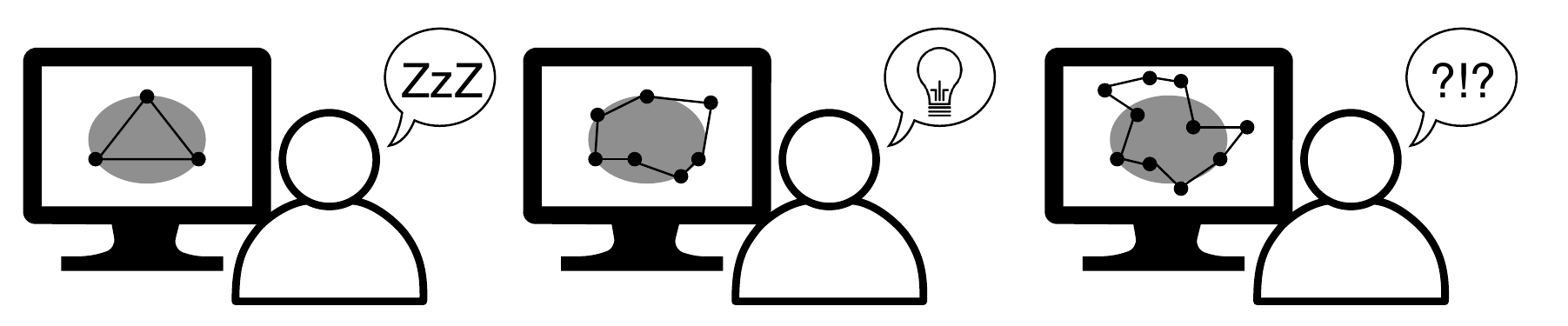} 
\caption{Various degrees of engagement with IML region digitization.  In the first image, the machine
has overshot cognitive load and thus the analyst's productivity is hampered.  In the second image,
the analyst is engaged in the task and the machine is helping their productivity.  In the last image,
the machine has undershot the cognitive load and thus the analyst is overwhelmed and will most likely
abandon the IML implementation for the task.}
\label{fig:analyst}
\end{figure*}

This example is simple, but it highlights one of the detrimental problems of IML implementations:
Overfitting is inevitable, and it can induce, rather than relieve, cognitive load.  As
mentioned previously, reinforcement learning may be used to augment the uncertainty or similarity
model based on the number of corrections the user has to make in any iteration.  However, 
convergence of such a technique would involve the user making excessive
corrections in order to
inform the model in this example.
Unlike AML, the uncertainty and workload involved with IML data must be somehow
informed by the analyst.

Figure \ref{fig:analyst} shows several situations exemplifying various
levels of cognitive load when an analyst uses GRAIT to annotate some region of interest.
In the first example, the machine is 
very accurate but offers too few vertices for the analyst to verify.  In this situation, the
analyst is impeded by an overshot cognitive load.  The analyst must work at the slow pace 
of the IML implementation, which not only reduces their productivity but may also reduce their
attention and engagement.  The second example shows the ideal situation where GRAIT
correctly manages the cognitive load of the analyst.  The analyst is expected to be engaged and
productive.  The last situation shows an example of the IML implementation undershooting the
cognitive load.  This causes the analyst to become overwhelmed and possibly confused, slowing
their productivity and causing frustration.

Incorporation of cognitive load is necessary to avoid the pitfall of bad cognitive load estimation
based on analysis of data alone. 
For instance, consider an augmentation to the third GRAIT example in the figure by 
providing the user with a survey at each iteration.  The survey will occur before correction
and simply ask, ``Is this workload too little, too much, or fine?''  In this particular situation,
the analyst will inform the machine that the workload is too much to handle, and the machine may
modify its uncertainty model accordingly (e.g. by adjusting weighting or performing best-fit
optimization to prior iterations).  This very simple solution illustrates how cognitive
feedback may enable better IML for many applications, but this concept may be taken further.  
In order to promote discussion and research of the possibilities and implications of this concept,
we now present a taxonomy for cognitive feedback to inform IML.

{\it Self-reported cognitive feedback} is gathered by surveys eliciting cognitive feedback from 
the user.  An example of such a survey is the standard NASA-TLX, which allows a user to report
on the general experienced workload of a particular task \cite{Hart1988}.
This could be gathered offline during human
factors evaluation or online through an interface for self reporting within the HCI.
The main advantage of online self reporting cognitive load
is the simplicity to collect feedback within the HCI.  
Implementation of simple interventions,
such as providing buttons for when a workload is too heavy or too light, are trivial.  However,
this approach may be imprecise in complex user environments because sub-components
of a task may differentially contribute to workload.  In these situations,
interventions may be too simplistic or induce load on an analyst.

Until now, we've discussed the implications of self reporting on cognitive load, but this
technique may provide insight into more than just the analyst's ideal workload.  The field of
explainable artificial intelligence involves expressing the machine's decision making
to a human user \cite{Gunning2019}.  If a model for explainability is feasible, then the user
may communicate cognitive information relating to features
as feedback to the model \cite{Teso2019}.  Relating back to the
example above, the machine may explain its decisions by stating ``I believe that historic
position of the shoreline is very important.''  The user may then augment the belief by
stating ``The historic position is not as important as color,'' and the machine may then
optimize its classifier and uncertainty calculation based on this statement. 

\begin{table*}[th]
\small
\centering
\caption{Taxonomy of Cognitive Feedback for Interactive Machine Learning}
\begin{tabular}{l l l}
\hline
Term & Definition & Examples\\
\hline
Self Reporting & Gathered by surveying the analyst.  & Online: Buttons in HCI.\\
               &                                     & Offline: Human-factors surveys.\\
Implicit       & Collection and evaluation of biofeedback & Cognitive load of correction via HCI.\\
               & via closed experimentation.              & Load as a function of correction count.\\
               &                                          & Use of cognitive cues as ML features.\\
Modeled        & Utilization of a cognitive model in & Feedback model of user interaction with HCI.\\
               & the loop.                           &\\
\hline
\end{tabular}
\label{tab:tax}
\end{table*}

As opposed to surveying a user, {\it implicit cognitive feedback} may be collected in
real time while analysts interact with the HCI during closed experimentation.  Implicit
cognitive feedback involves collecting physiological data in order to infer cognitive
states in a manner that is continuous, objective, and occurs in real time.  
For example, because pupillary responses are reflective of nervous activity, pupil
dilation may act as a proxy for measuring task-induced cognitive processes.  As such,
increases in pupil diameter may be indicative of high cognitive load, attentional
processing, and decision making \cite{hess1964pupil,kahneman1973attention,hahnemann1967pupillary}
whereas decreases may reflect fatigue \cite{lowenstein1963pupillary}.
This data
may then be correlated with self-reporting to define various states of cognitive load.
Examples of such biofeedback include readings of skin conductance, heart rate,
pupilometry, and electroencephalogram (EEG).  Often, multiple physiological measures
will be assessed to determine workload and inform adaptive algorithms, in essence
creating user models that dynamically adjust to support user needs.  For example,
such physiological elements were examined to monitor the workload of operators while
performing UAV piloting tasks of different levels \cite{Wilson2007}.  The physiological
signals were used as features to train a neural network to classify workload.
Another approach of implicit cognitive feedback is to incorporate
cognitive cues as features in the 
machine learning algorithm \cite{rosenfeld_combining_2012}. For example, in a 
recent choice competition, researchers incorporated cognitive features derived 
from behavior into a random forest algorithm. They found that this approach 
significantly outperformed other ML approaches that did not incorporate 
cognitive features \cite{plonsky_psychological_2017}. 
A recent study has explored how collecting and applying cognitive cues as features
improves reinforcement learning algorithms for playing video games \cite{Zhang2019}.
In summary, implicit cognitive feedback has the potential to improve IML implementations
by gathering data in closed experimentation to inform cognitive load, uncertainty/similarity
measurements, and inform the machine with features of interest related to a specific task.

Implicit cognitive feedback may provide invaluable insight to IML implementations, but the
disadvantage lies in the fact that closed experimentation is often necessary to collect
biofeedback, control levels of tasking, and survey users of the HCI with respect to a particular
application.  Additionally, the cognitive state of the user may be more dynamic for some
applications than others.  In these situations, {\it modeled cognitive feedback} may provide
cognitive feedback based on models of user interaction with the HCI.
For example, simulating human behavior using a computational cognitive model is another 
potential method to provide feedback to an IML system. 
Models of cognition and decision making have been used to simulate human interactions
with interfaces in military contexts \cite{blasch2011user}.
Cognitive architectures represent a modeling paradigm that
computationally defines the relationship between underlying biological and 
cognitive mechanisms to emerging behavior. Architectures, such as ACT-R 
\cite{anderson_integrated_2004} and SOAR \cite{laird_soar_1987}, have long 
been a part of HCI research to simulate users interacting with an interface. 
For example, ACT-R models are used for usability testing of menus 
\cite{byrne_act-r/pm_2001}, modeling how users detect phishing websites 
\cite{williams_simulating_2017}, and detecting situations with high cognitive 
load when using a smartphone \cite{wirzberger_modeling_2015}. Cognitive 
architectures have been used with physiological data, such as eye tracking 
information and fMRI, to map observed behavior the underlying mental states 
and brain regions  \cite{tamborello_adaptive_2007,borst_using_2015}. Cognitive 
models, combined with self-reported data from surveys and physiological data, 
can provide a starting point for IML systems to optimize their suggestions for 
the overall performance of a human-machine team.

These three different categories of cognitive feedback -- self reporting, implicit cognitive
feedback, and modeled cognitive feedback -- delineate the possible ways in which IML 
implementations may 
be centered around the analyst.  The categories are summarized in Table~\ref{tab:tax}.

Once cognitive feedback has been integrated into IML, more conventional results
such as classification accuracy and overall corrections may be used to evaluate
approaches against their non-cognitive baseline.
However, these results may lack true insight into the purpose of the human-machine
team.  Measuring the cognitive
load on human subjects with more objective metrics of productivity would provide
more insight into the effectiveness of IML implementations \cite{Alves2016}.
Additionally, it is the analyst themselves who must also evaluate the effectiveness
of an IML implementation, though this may
take high levels of time and effort \cite{Groce2013,Gillies2016}.  

\section{A Future Driven by Cognitive Feedback}

We have presented a summary of interactive machine learning along with several examples
informing the state of the art.  After discussing the advantages of IML, the major 
shortcomings and gaps were delineated.  Finally, the implications of cognitive feedback
for IML implementations were discussed to address the gaps.
Though it may seem trivial to study cognitive feedback as it relates to data science
for human-in-the-loop applications, there is a general lack of such studies in the literature,
especially for defense applications.  We hope this article will encourage research and
development in more IML for defense applications and more research in how cognitive 
feedback may inform IML implementations.

\bibliography{cog,vision,cogmodels}
\bibliographystyle{apalike}

\end{document}